# Color theorems, chiral domain topology and magnetic properties of $Fe_xTaS_2$

Yoichi Horibe[1+], Junjie Yang[2,3], Yong-Heum Cho[2,3], Xuan Luo[2,3], Sung Baek Kim[2,3#], Yoon Seok Oh[1], Fei-Ting Huang[1], Toshihiro Asada[4], Makoto Tanimura[4], Dalyoung Jeong[5], and Sang-Wook Cheong[1*],

[1]*Rutgers Center for Emergent Materials and Department of Physics & Astronomy, Rutgers University, Piscataway, New Jersey 08854*

[2]*Laboratory for Pohang Emergent Materials, Pohang University of Science and Technology, Pohang 790-784, Korea*

[3]*Department of Physics, Pohang University of Science and Technology, Pohang 790-784, Korea*

[4]*Research Department, Nissan Arc, Ltd., Yokosuka, Kanagawa 237-0061, Japan*

[5]*Department of Mathematics, Soongsil University, Seoul 156-743, Korea*

**ABSTRACT:**

**Common mathematical theory can have profound applications in understanding real materials. The intrinsic connection between aperiodic orders observed in the Fibonacci sequence, Penrose tiling, and quasicrystals is a well-known example. Another example is the self-similarity in fractals and dendrites. From transmission electron microscopy experiments, we found that $Fe_xTaS_2$ crystals with x=1/4 and 1/3 exhibit complicated antiphase and chiral domain structures related to ordering of intercalated Fe ions with 2a×2a and √3a×√3a superstructures, respectively. These complex domain patterns are found to be deeply related with the four color theorem, stating that four colors are sufficient to identify the countries on a planar map with proper coloring, and its variations**

**for two-step proper coloring. Furthermore, the domain topology is closely relevant to their magnetic properties. Our discovery unveils the importance of understanding the global topology of domain configurations in functional materials.**

■ INTRODUCTION

Understanding and controlling domains and domain walls is quintessential for identifying the origin of the macroscopic physical properties of functional materials and exploiting them for technological applications. Domains are associated with different orientations of directional order parameters such as magnetization, polarization, and ferroelastic distortions[1-3]. Even though the local conditions at, for example, ferroelastic boundaries and liquid crystal defects have been often studied[4,5], research on the macroscopic topological constraints in complex domain patterns has been scarce.

Layered transition metal dichalcogenides, consisting of $MC_2$ layers (M: transition metal elements, C: chalcogen elements), exhibit fascinating physical properties such as charge-density waves (CDWs)[6], superconductivity[6-9], and discommensurations[10-15]. Intercalation of other transition metal ions between the $MC_2$ layers gives rise to distinct superstructures, leading to significant changes in crystallographic structures and physical properties. Fe-intercalated $TaS_2$ shows highly anisotropic ferromagnetism at low temperatures[16-20]. Ordering of the intercalated Fe ions in $Fe_xTaS_2$ with x=1/4 and 1/3 results in 2a×2a and √3a×√3a superstructures (a: the hexagonal lattice parameter of non-intercalated $TaS_2$) with the space groups of centrosymmetric *P6₃/mmc* and noncentrosymmetric (chiral) *P6₃22*, respectively[16,17]. These 2a×2a and √3a×√3a superstructures are accompanied by four types (A, B, C, and D) and three types (A, B, and C) of structural antiphase domains in each intercalated Fe plane of $Fe_{1/4}TaS_2$ and $Fe_{1/3}TaS_2$, respectively. These antiphase domains are associated with the different origins (i.e., phase shifts) of the superstructures.

Herein, we report that the global topology of the antiphase and chiral domain patterns in $Fe_xTaS_2$ appears to be complex, but can be neatly understood in terms of the four color theorem[21-23] as well as a tensorial color theorem associated with two-step proper coloring. The four color theorem, which was empirically known to cartographers before the 17th century, states that four colors are sufficient to identify the countries on a planar map with proper coloring (without bordering countries sharing the same color, except for intersections). The tensorial coloring is found to be related to the presence of chiral and antiphase domains in $Fe_{1/3}TaS_2$. Intriguingly, the global domain patterns associated with color theorems are directly relevant to bulk magnetic properties in $Fe_xTaS_2$. The relevance of color theorems to the domain patterns and physical properties of $Fe_xTaS_2$ resembles the relation between the order without periodicity in the Fibonacci sequence, Penrose tiling and the formation of quasicrystals[24-26]; it is also similar to the way that self-similarity plays the key role in the formation of fractals and dendrites[27]. Our discovery unveils the importance of understanding the global topology of domain configurations in functional materials.

■ EXPERIMENTAL SECTION

Thin plate-like single crystals of $Fe_xTaS_2$ were grown by a chemical vapor transport method[19,20,28]. Magnetic hysteresis curves of the crystals were obtained using a Quantum Design Magnetic Property Measurement System, and the real Fe compositions were estimated from the saturation magnetic moments in the magnetic hysteresis curves with the assumption that each mole of $Fe^{2+}$ ions exhibits the full moment of 4 $\mu_B$/mole. The estimated Fe compositions of crystals with nominal compositions of 0.33, 0.4, 0.5, 0.6, and 0.7 are 0.18, 0.25, 0.34, 0.43, and 0.47, respectively. The estimated concentrations were used for the x values in this paper, and the samples with x=0.25, 0.34, 0.43, and 0.47 were used for transmission electron microscopy (TEM) studies to investigate the superlattice domain structures. Specimens for TEM studies were prepared by mechanical cleaving, followed by Ar ion milling. TEM studies were carried out using JEOL-2010F and JEOL-2000FX microscopes at room temperature. The electron diffraction patterns and real-space images were recorded with a 14-bit charge-coupled-device array detector and imaging plates. The reflection spots in the diffraction patterns were indexed using a simple hexagonal notation.

## ■ RESULTS

The results of our TEM experiments reveal that in single-crystalline $Fe_xTaS_2$, $Fe_{0.25}$ shows 2a×2a-type superlattice reflections, while $Fe_{0.34}$, $Fe_{0.43}$, and $Fe_{0.47}$ exhibit the √3a×√3a type, indicating that the change in the superstructure occurs somewhere between x=1/4 and 1/3. Figures 1a and 1e display electron diffraction patterns of $Fe_{0.25}$ and $Fe_{0.43}$, respectively. The superlattice reflection spots are observed at the (1/2 0 0)-type and (1/3 1/3 0)-type reciprocal positions in $Fe_{0.25}$ and $Fe_{0.43}$, indicating the presence of 2a×2a and √3a×√3a-type superstructures, respectively. Note that the intercalated Fe ions form a two-dimensional (2D) supercell between two adjacent Ta-$S_6$ prismatic layers (Supplementary Information, section S1). The distinct feature between the 2a×2a and √3a×√3a superstructures in $Fe_xTaS_2$ is the different stacking sequence of the 2D supercells along the c-axis. Specifically, the 2a×2a superstructure consists of identically stacked 2D supercells (i.e., AA-type stacking), while the √3a×√3a superstructure contains shifted 2D supercells with AB-type stacking, as shown in Fig. 1b and Fig. 1f, respectively. These different stacking sequences result in the centrosymmetric *P6$_3$/mmc* and noncentrosymmetric and chiral *P6$_3$22* space groups for the 2a×2a and √3a×√3a superstructures, respectively.

We found complicated configurations of antiphase domains in the dark-field images of Fe$_{0.25}$ and Fe$_{0.43}$ taken using superlattice spots. The antiphase domain patterns of two specimens exhibit distinct features, even though their diffraction patterns appear similar, only differing in the exact position of the superlattice spots. For example, the antiphase domains in Fe$_{0.25}$ are significantly smaller than those in Fe$_{0.43}$. Figures 1c and 1g show the dark-field images taken using the ($\bar{1}$/2 0 0) superlattice spot indicated as S1 in Fig. 1a and the ($\bar{1}$/3 $\bar{1}$/3 0) superlattice spot denoted as S4 in Fig. 1e, respectively. The antiphase boundaries, separating neighboring antiphase domains, are clearly visible as dark line contrasts in superlattice dark-field images in both cases. (Supplementary Information, section S2).

There is an extinction rule for the dark-field images of antiphase boundaries in the 2a×2a superstructure. For example, the antiphase boundary between the BB-type and CC-type antiphase domains appears in the S1=($\bar{1}$/2 00) (Fig 2a) and S2=(0 1/2 0) (Fig. 2b) dark-field images, but disappears in the S3=(1/2 $\bar{1}$/2 0) dark-field image of Fig. 2c (see also Supplementary Information, section S3). Each antiphase boundary becomes invisible in a dark-field image taken using one out of three superlattice spots (namely, S1, S2, or S3), when no antiphase shift at the boundary exists along a certain superlattice modulation wave vector. This absence of antiphase shifts at the antiphase boundary leads to the extinction rule for the antiphase boundaries in superlattice dark-field images. This rule is summarized in Fig. 3, showing the local structures near boundaries between two antiphase domains. The boundaries are highlighted with yellow bands, and the three directions of the superlattice modulation wave vectors are denoted by S1, S2, and S3, respectively, as shown in Fig. 1a. The red, yellow, blue, and green circles correspond to AA-, BB-, CC-, and DD-type superstructures, respectively, which are associated with four possible origins of the 2a×2a Fe superstructure. It is evident that the superlattice modulation along only one out of three equivalent crystallographic directions does not show any antiphase shift; this is indicated by light green dashed lines (along the S1 direction), light blue dashed lines (along the S2 direction), and pink dashed lines (along the S3 direction). For example, the antiphase boundary between BB-type and CC-type (or AA-type and DD-type) antiphase domains has antiphase shifts along the S1 and S2 directions, while no antiphase shift along the S3 direction. In other words, when a boundary along a BB-type (AA-type) antiphase domain disappears in a S3 superlattice dark-field image, the neighboring domain should be CC-type (DD-type). This extinction rule of antiphase boundaries allows us to identify four different antiphase domains in any complicated domain patterns. Furthermore, the exact nature of the antiphase boundaries in the boxed area in Figures 2a-2c is unveiled in high-resolution TEM images (Figures 2d-2f), showing the lattice fringes of superlattice modulations. The specimen was tilted along the 100-type reciprocal axes to excite the superlattice reflection spots for Figs. 2a-2c, so the lattice fringes are perpendicular to the superlattice modulation wave vectors $\vec{q}$. The high-resolution TEM images clearly demonstrate that antiphase shifts appear only in Fig. 2a and 2b, but there exists no antiphase shift in Fig. 2c, indicating that this extinction rule originates from the local structure near the antiphase boundary (Supplementary Information, section S4).

This extinction rule for the antiphase boundaries allows us to identify the four different antiphase domains in a complicated 2a×2a superstructure domain configuration. A complete map of antiphase boundaries and domains obtained from all three distinct dark-field images (taken using three different superlattice spots, i.e., S1, S2, and S3) for the area in the red box in Fig. 1c is shown in Fig. 1d. We found that the complex antiphase domain pattern of the 2a×2a superstructure in Fig. 1c can be neatly understood in terms of the four color theorem.

Mathematically, the four color theorem is stated as "all faces of every planar graph can be four-proper-colorable or $Z_4$-colorable". Consistent with this four color theorem, the pattern in Fig. 1d for the 2a×2a superstructure is $Z_4$ (red, blue, yellow, and green)-colorable, and these $Z_4$ colors correspond to the AA-, BB-, CC-, and DD-type 2a×2a superstructures, respectively. The physical meaning of this intriguing correspondence between the four color theorem and the pattern for the 2a×2a superstructure can be interpreted as follows: [1] there exists no particular rule for the 2a×2a superstructure domain pattern, [2] domain size tends to be small if there exists no rule for a global domain configuration, and [3] most importantly, all boundaries are coherent, and there exist no disordered (incoherent) boundaries in the 2a×2a superstructure domain pattern.

In contrast to that of the 2a×2a superstructure, the domain configuration of the √3a×√3a superstructure exhibits a unique topology, and the domains tend to be significantly larger. First of all, the typical domain size in $Fe_{0.43}$ is about 3 μm, while that in $Fe_{0.25}$ is about 150 nm. Second, six antiphase boundaries of the √3a×√3a superstructure merge always at one point without any exception, and thus the superstructure domain pattern forms a 6-valent graph. Furthermore, each domain is always surrounded by an even number of vertices, thus forming a so-called "even-gon". The concept of the standard one-step proper coloring can be extended to that of two-step (tensorial) proper coloring, and all faces of every 6-valent graph with even-gons are tensorial-$Z_2 \times Z_3$-colorable[29]. It turns out that all faces of every 6-valent graph with even-gons are two-proper-colorable. After this 1st-step two-proper coloring (with dark and light colors), faces with dark or light colors can be further colored with three colors (with red, blue, and green colors) in such a way that each face is surrounded neither by any face with the same "first" color nor by any face with the same "second" color. For example, a dark red face is never surrounded by any light red face or any dark faces. This non-trivial process turns out to be unique. This $Z_2 \times Z_3$-coloring for the area in the blue box in Fig. 1g is displayed in Fig. 1h (see also Supplementary Information, section S5).

The $Z_4$-coloring and $Z_2 \times Z_3$ tensorial coloring in $Fe_{0.25}$ and $Fe_{0.43}$, respectively, is summarized in Fig. 4. Figure 4a shows the superlattice dark-field image of $Fe_{0.25}$ from the red-framed area in Fig. 1c. The antiphase domain boundaries can be observed as dark lines. A complete antiphase domain pattern of $Fe_{0.25}$ with $Z_4$ coloring is depicted in Fig. 4b. Note that the domain pattern is obtained from superlattice dark-field images taken using various superlattice spots. All antiphase domains were identified using the extinction rule and are denoted AA, BB, CC, and DD; these four types of antiphase domains are represented by $Z_4$ colors (red, blue, green, and yellow). Figures 4c and 4d show the superlattice dark-field image of $Fe_{0.43}$ (from the blue-framed area in Fig. 1g) and an antiphase and chiral domain pattern of $Fe_{0.43}$ with $Z_2 \times Z_3$ coloring. The $Z_2 \times Z_3$ colors have one-to-one correspondence with chiral and antiphase domains; in other words, $Z_2$ corresponds to two types of chirality, and $Z_3$ represents three antiphase domains.

The $Z_2 \times Z_3$ coloring for the $\sqrt{3}a \times \sqrt{3}a$ superstructure domains is associated with profound physical consequences. First, $Z_3$ (red, blue, and green) colors correspond to three types of $\sqrt{3}a \times \sqrt{3}a$ superstructure antiphase domains, i.e., AB- , BC-, and CA-type antiphase domains. The $Z_2$ coloring is associated with chiral domains without centrosymmetry. These chiral domains are clearly observed in the dark-field image taken using the S4=($\bar{1}/3$ $\bar{1}/3$ 0) superlattice spot under the so-called Friedel's-Pair-Breaking condition[30], as shown in Figure 2h. On the other hand, the dark-field image in Figure 2g, taken using the same S4 superlattice spot with the electron incidence almost parallel to the [001] zone axis, shows only antiphase boundaries. The top-view and side-view stacking structures of the 2D Fe supercells in $x \approx 1/4$ (AA-type) and in $x \approx 1/3$ (AB-type) are depicted in Figs. 5a and 5b, respectively. Inversion symmetry is broken in $x \approx 1/3$ because the inversion center of the TaS$_2$ lattice along the c axis does not match with that of the Fe lattice with the AB-type stacking, while inversion symmetry remains intact in x≈1/4 with the AA-type stacking. Intercalation of Fe ions tends to induce rotation of Ta-S$_6$ prisms around Fe ions in x≈1/3, and the opposite rotation of the top and bottom Ta-S$_6$ prisms around each intercalated Fe ion induces chirality in x≈1/3 with the AB-type stacking (Supplementary Information, section S1 and S6).

■ DISCUSSION

The diffusion of intercalated Fe ions is expected to be important for the formation of the complex domains in Fe$_x$TaS$_2$, and the stacking of 2D supercells (AA-type in x≈1/4 or AB-type in x≈1/3) may play a key role in forming the $Z_4$ or $Z_2 \times Z_3$ domain patterns. In x≈1/4, the intercalated Fe ions sit at the same position along the c-axis and the Ta ions in between are therefore not distorted (Fig. 5a). As a result, four types of antiphase domains related to the presence of four Fe positions (i.e., AA-, BB-, CC-, and DD-type) can form during the disorder-order transition of Fe ions, resulting in the $Z_4$ domain pattern where the antiphase domains are distributed without any regularity. On the other hand, in x≈1/3, the intercalated Fe ions on the upper and lower Fe layers are shifted and Ta ions can therefore distort along the c-axis (Fig. 5b). These Ta distortions can induce intriguing energetics in such a way that, for example, near the boundary region between A- and B-type 2D supercells within one layer, a C-type supercell is stable in the next layer, which leads to the two adjacent AC- and BC-type domains (Supplementary Information, section S4). A topological defect structure with six BC/AC/AB/CB/CA/BA domains (in a counterclockwise direction) can be realized through this process, as shown in Fig. 5c. Note that [1] when cyclicity and chirality is considered, BC/AC/AB/CB/CA/BA domains can be renamed A+/B-/C+/A-/B+/C- domains, respectively (Supplementary Information, section S7), [2] a vorticity can be defined for a topological defect with A+/B-/C+/A-/B+/C- domains, so the defect can be considered a topological vortex with $Z_2 \times Z_3$ patterns, i.e., a $Z_2 \times Z_3$ vortex, and [3] an adjacent topological defect with A+/C-/B+/A-/C+/B- has the opposite vorticity, so it can be considered a topological antivortex. Vortices (blue circles) and antivortices (red circles) associated with the antiphase and chiral domains are identified in Fig. 4e. It is evident that a vortex is surrounded by antivortices and vice versa.

We found that the domain topology in $Fe_xTaS_2$ has a significant effect on the magnetic properties. Figure 5d shows magnetic hysteresis (M(H)) curves measured at 4 K with the magnetic field parallel to the c axis, and magnetic coercivity, extracted from M(H) curves, vs. x in $Fe_xTaS_2$. The M(H) curves of $Fe_{0.18}$ and $Fe_{0.25}$, showing $Z_4$ domain patterns with small domain sizes, exhibit sharp switching of magnetization with large coercivities. On the other hand, the M(H) curves of $Fe_{0.34}$ and $Fe_{0.47}$ with $Z_2 \times Z_3$ domain patterns with large domain sizes show relatively broad switching of magnetization with small coercivity (Supplementary Information, section S8). High-density antiphase boundaries associated with $Z_4$ domains are probably associated with a strong pinning effect of magnetic domain walls and lead to an avalanche-like depinning of magnetic domain walls in the presence of high external magnetic fields[19, 20]. On the other hand, large $Z_2 \times Z_3$ domains with a small number of antiphase/chiral domain boundaries accompany weak pinning of magnetic domain walls and a small coercivity field.

In summary, Fe ionic ordering in $Fe_xTaS_2$ induces two types of superstructures: centrosymmetric 2a×2a and chiral √3a×√3a types. The topologies of the complex antiphase domains of these 2a×2a and √3a×√3a superstructures are associated with $Z_4$ and $Z_2 \times Z_3$ coloring, respectively. We also found that these different topologies have significant physical consequences, such as domain size and magnetic properties. Our findings provide new insights for understanding and controlling domains in complex functional materials.

■ ASSOCIATED CONTENT

Supporting Information. Description of the material included. This material is available free of charge via the Internet at http://pubs.acs.org.

■ AUTHOR INFORMATION

Corresponding Author

sangc@physics.rutgers.edu

[+]Current address: Department of Materials Sciences and Engineering, Kyushu Institute of Technology, Fukuoka 804-8550, Japan

[#]Current address: Advancement for College Education Center, Konyang University, Chungnam 320-711, Korea.

The authors declare no competing financial interests. Correspondence and requests for materials should be addressed to S.-W.C (sangc@physics.rutgers.edu).

■ ACKNOWLEDGEMENTS

We would like to thank Y. J. Choi and Weida Wu (Rutgers) for useful discussion. This work was supported by the NSF under Grant No. NSF-DMREF-1104484. The work at Postech was supported by the Max Planck POSTECH/KOREA Research Initiative Program [Grant No. 2011-0031558] through NRF of Korea funded by MEST.

Figure Captions

Fig. 1.

Electron diffraction patterns and domain structures of $Fe_{1/4}TaS_2$ and $Fe_{1/3}TaS_2$. **a,e,** Electron diffraction patterns of $Fe_xTaS_2$ with x≈1/4 and 1/3, respectively The 2a×2a -type (indicated by S1, S2 and S3) and√3a×√3a -type (indicated by S4) superlattice spots can be observed clearly in addition to the fundamental spots. Note that the electron incidence is almost parallel to the [001] direction. **b,f,** Two dimensional schematics of the 2a×2a and √3a×√3a superstructures of intercalated Fe ions in $Fe_{1/4}TaS_2$ and $Fe_{1/3}TaS_2$, respectively. The red and blue spheres depict Fe ions, and small (large) spheres represent the lower (upper) Fe layers. **c,g,** Dark-field images taken using the superlattice spots indicated by S1 and S4, respectively. Note that some antiphase boundaries are invisible in **c** because of the extinction rule. **d,h,** Domain patterns with four proper coloring (red, yellow, blue, and green) in $Fe_{1/4}TaS_2$ and tensorial proper coloring (1st step: dark and light, 2nd step: red, blue, and green) in $Fe_{1/3}TaS_2$, respectively. The **d** and **h** schematics correspond to the red- and blue-framed areas in **c** and **g**, respectively.

Fig. 2.

Antiphase and chiral domains in $Fe_xTaS_2$. **a-c,** Dark-field images of $Fe_{0.25}TaS_2$ taken using superlattice spots indicated by S1, S2, and S3 in Fig. 1**a**, respectively. These images correspond to the black-box area in Fig. 1**d**. **d-f,** High-resolution TEM images from the color-box areas in **a**, **b**, and **c**, respectively. The antiphase boundary is indicated with a dashed yellow line. **g,h,** Dark-field images of $Fe_{0.43}TaS_2$ taken using the superlattice spot indicated by S4 in Fig. 1**e**. The dark-field image in **h** taken under the so-called Friedel's-pair-breaking condition clearly exhibits the presence of chiral domains without centrosymmetry.

Fig. 3.

Extinction rule and local structure of antiphase boundaries in $Fe_{1/4}TaS_2$. **a,** A schematic of the [001] electron diffraction pattern with S1, S2, and S3 superlattice modulation wave vectors. **b,** Four possible origins of the 2a×2a Fe superstructure. The red, yellow, blue, and green circles correspond to AA-, BB-, CC-, and DD-type superstructures, respectively. The AA-type 2a×2a superstructure is depicted with thick red lines. **c,** Local structures near boundaries between two antiphase domains. The boundaries are highlighted with yellow bands. The superlattice modulation along one out of three equivalent crystallographic directions does not show any antiphase shift, resulting in the absence of the antiphase boundary contrast in a superlattice dark-field image.

Fig. 4.

Antiphase domains in $Fe_xTaS_2$ and proper coloring of them. **a,** Superlattice dark-field image of $Fe_{0.25}$ from the red-framed area in Fig. 1c. Note that some antiphase boundaries are invisible because of the extinction rule. **b,** A complete antiphase domain pattern of $Fe_{0.25}$ with $Z_4$ coloring. All antiphase domains were identified using the extinction rule and are consistent with the $Z_4$ colors. **c,** superlattice dark-field image of $Fe_{0.43}$ from the blue-framed area in Fig. 1g. **d,** An antiphase and chiral domain pattern of $Fe_{0.43}$ with $Z_2 \times Z_3$ coloring. **e,** $Z_2 \times Z_3$ vortices and antivortices in the antiphase and chiral domain pattern. The vortices and antivortices are indicated by blue and red circles, respectively.

Fig. 5.

Domain evolution and magnetic properties in $Fe_xTaS_2$. **a,b,** The top-view and side-view schematics of the crystallographic structures of $Fe_{1/4}TaS_2$ and $Fe_{1/3}TaS_2$, respectively. Only Fe (red) and Ta (green) ions are depicted. The side-view schematics correspond to the portions indicated in orange in the top-view schematics. The arrows depict the displacement of Ta ions along the c axis. **c,** Schematics of the evolution of a $Z_2 \times Z_3$ domain during the disorder-order transition of Fe ions. **d,** Magnetic hysteresis curves of x=0.18, 0.25, 0.34, and 0.47. These curves were measured at 4 K in magnetic fields along the c axis. Note that x=0.18 and 0.25 show the 2a×2a-type superstructure, while x=0.34 and 0.47 exhibit the √3a×√3a-type one. The inset shows the magnetic coercivity as a function of Fe composition.

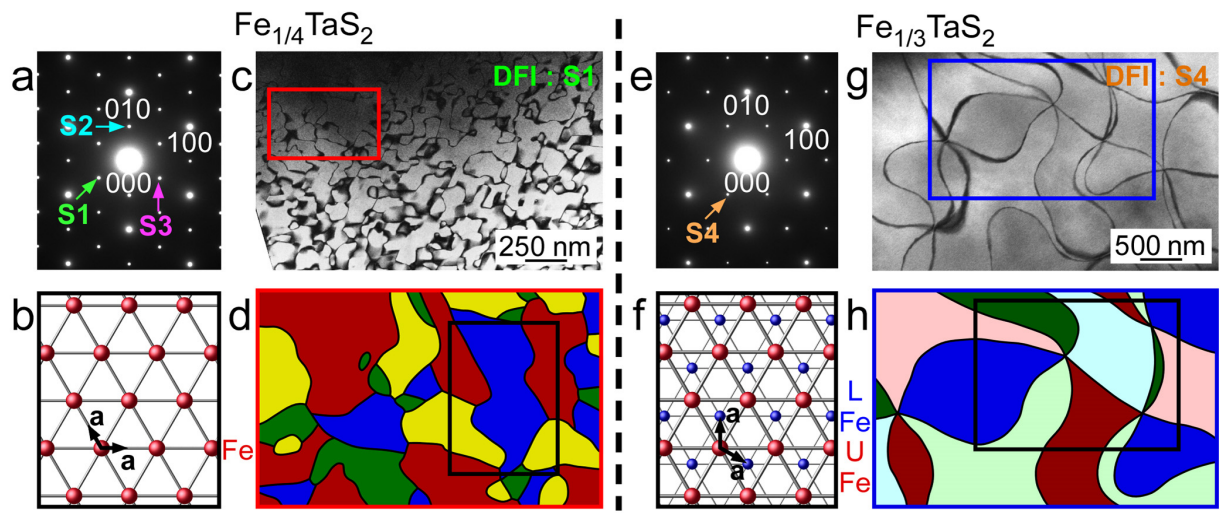

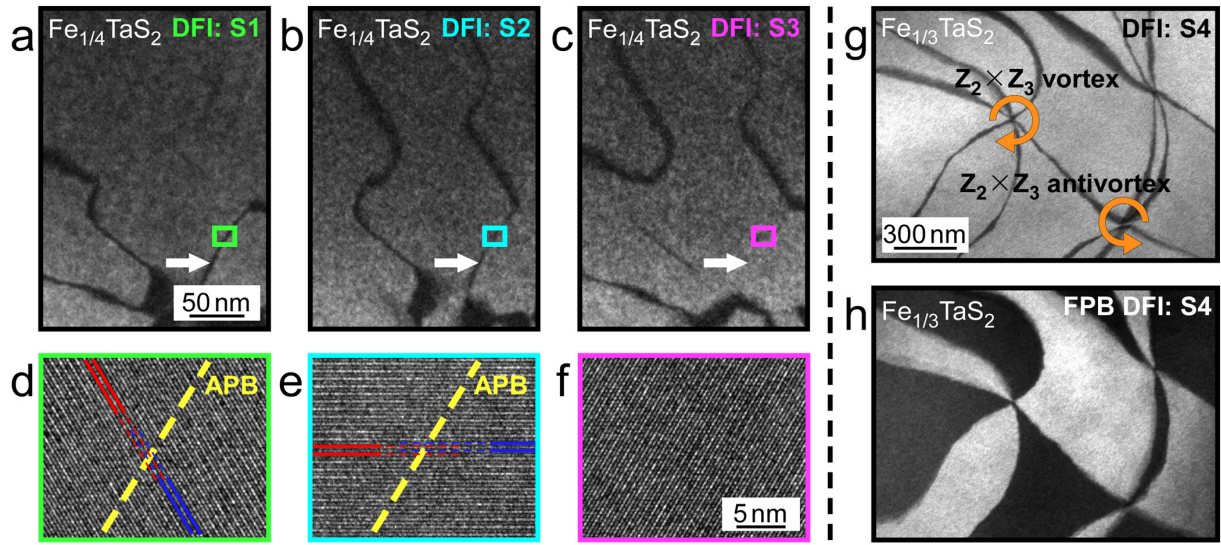

Fig. 2. Y. Horibe *et al.*,

- 19 -

Fig. 3.     Y. Horibe *et al.*,

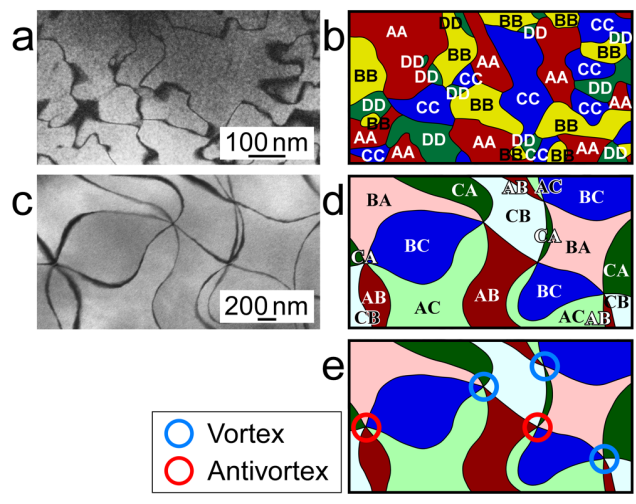

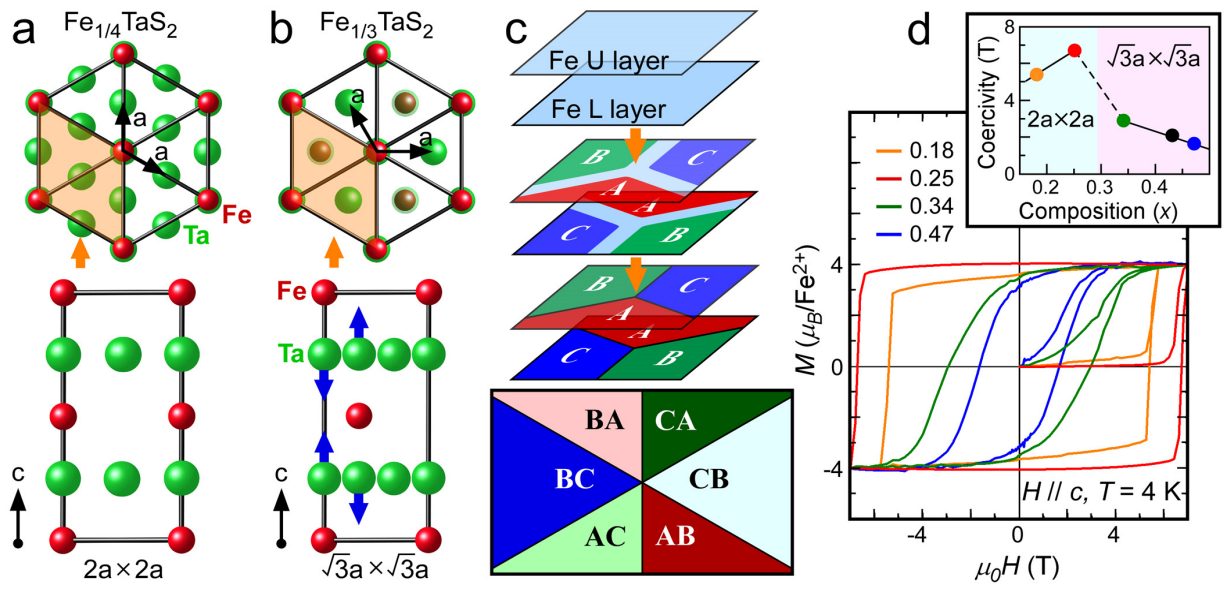

Fig. 5. Y. Horibe *et al.*,

Insert Table of Contents artwork here

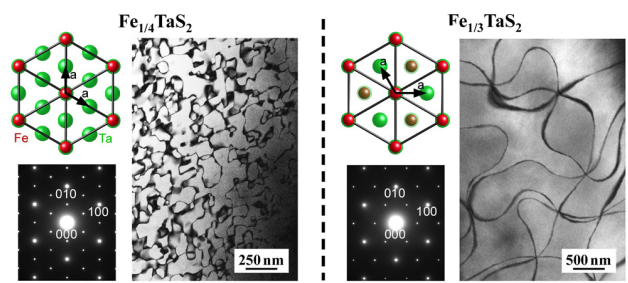

# Color theorems, chiral domain topology and magnetic properties of $Fe_xTaS_2$

Yoichi Horibe[1], Junjie Yang[2,3], Yong-Heum Cho[2,3], Xuan Luo[2,3], Sung Baek Kim[2,3+], Yoon Seok Oh[1], Fei-Ting Huang[1], Toshihiro Asada[4], Makoto Tanimura[4], Dalyoung Jeong[5], and Sang-Wook Cheong[1*]

[1]*Rutgers Center for Emergent Materials and Department of Physics & Astronomy, Rutgers University, Piscataway, New Jersey 08854*

[2]*Laboratory for Pohang Emergent Materials, Pohang University of Science and Technology, Pohang 790-784, Korea*

[3]*Department of Physics, Pohang University of Science and Technology, Pohang 790-784, Korea*

[4]*Research Department, Nissan Arc, Ltd., Yokosuka, Kanagawa 237-0061, Japan*

[5]*Department of Mathematics, Soongsil University, Seoul 156-743, Korea*

*email: sangc@physics.rutgers.edu

Supplementary information

### Section S1. Crystal structures of $Fe_{1/4}TaS_2$ and $Fe_{1/3}TaS_2$

Figures S1a and S1b display the three-dimensional (3D) crystallographic structures of $Fe_{1/4}TaS_2$ and $Fe_{1/3}TaS_2$, respectively. The red, green, and yellow spheres represent Fe, Ta, and S ions, respectively. In $Fe_xTaS_2$, the Ta ions occupy the centers of S prisms forming $Ta-S_6$ prismatic layers, and the intercalated Fe ions form a 2a×2a- or √3a×√3a-type 2D supercell within a Fe layer between two $Ta-S_6$ prismatic layers. The distinct feature between 3D 2a×2a and √3a×√3a superstructures of $Fe_xTaS_2$ is the stacking sequence of the supercells of the intercalated Fe ions along the c-axis. Namely, the 2a×2a superstructure consists of the identical stacking of the 2D Fe supercells ions along the c-axis (i.e., AA-type stacking), while the √3a×√3a superstructure contains an alternating stacking of the 2D Fe supercells (i.e., AB-type stacking). The presence of the 2a×2a and √3a×√3a superlattice reflections is consistent with the crystal structures with the hexagonal space group of *P6_3/mmc* and *P6_322*, respectively[1,2]. Note that in $Fe_{1/3}TaS_2$, the centrosymmetry is broken due to the coexistence of the AB-type stacking of the 2D Fe superstructure and the Ta-$S_6$ prismatic layers, while the crystal structure of $Fe_{1/4}TaS_2$ with the AA-type stacking of the 2D Fe superstructure is centrosymmetric.

A characteristic feature in the crystal structure of $Fe_{1/3}TaS_2$ is the rotation of Ta-$S_6$ prisms (Fig. S1c), leading to the presence of structural chirality. The expansion of neighboring S triangles of an intercalated Fe ion and the AB-type stacking of the 2D Fe supercells lead to the rotation of the upper and lower Ta-$S_6$ prisms around the Fe ion in the opposite directions, as indicated by light-blue and light-green arrows in Fig. S1d.

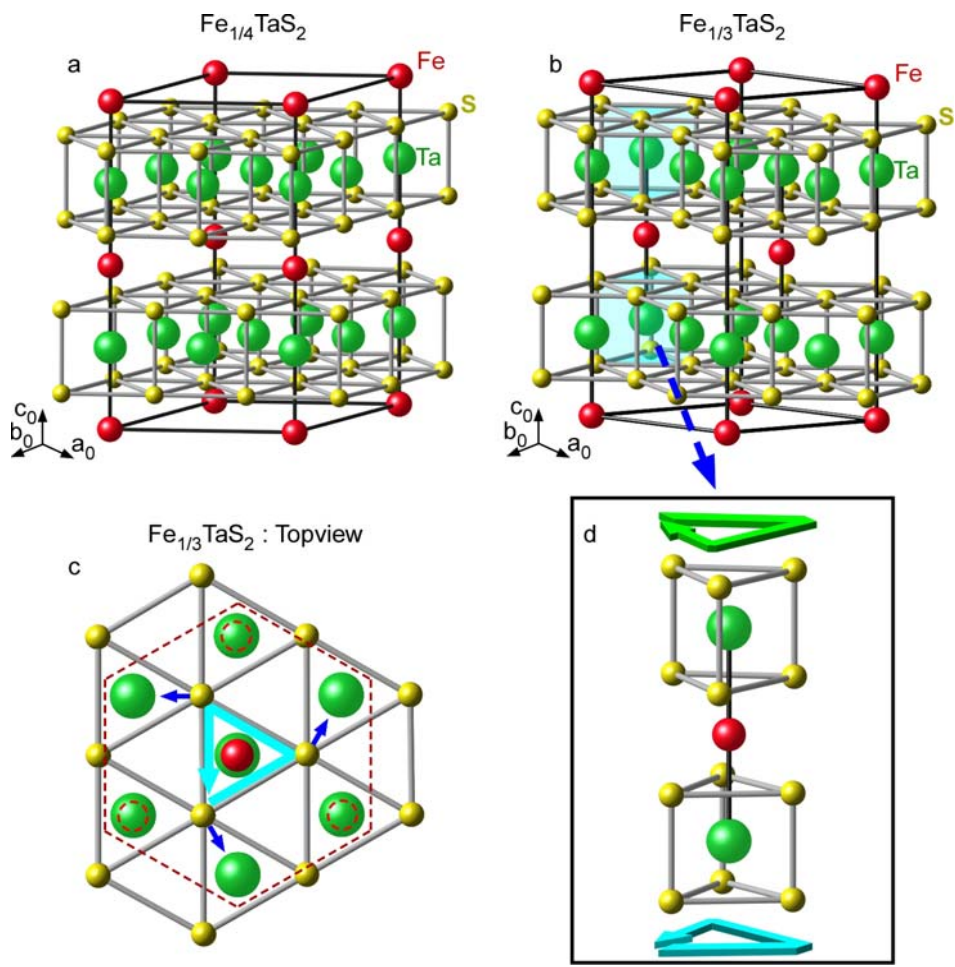

Fig. S1. Crystallographic structures of $Fe_xTaS_2$ with **a** $x \approx 1/4$ and **b** $x \approx 1/3$, respectively. The red, green, and yellow spheres represent Fe, Ta, S ions, respectively. **c,** The rotation of the Ta-$S_6$ prism below an intercalated Fe ion in the upper layer. The circles with dashed red-line represent the Fe ions in the bottom

layer. The chirality is associated with the rotation of a Ta-S$_6$ prism. **d,** The opposite rotations of two Ta-S$_6$ prisms locating above and below an intercalated Fe ion are depicted with light blue and green arrows.

**Section S2. Antiphase domain boundaries in Fe$_{1/4}$TaS$_2$ and Fe$_{1/3}$TaS$_2$**

To confirm the presence of antiphase boundaries associated with the 2a×2a and √3a×√3a superstructures, we performed dark-field imaging experiments using both fundamental and superlattice reflection spots in Fe$_{0.25}$ and Fe$_{0.43}$. Figure S2-1a displays an electron diffraction pattern observed in Fe$_{0.25}$, and shows (1/2 0 0)-type superlattice reflection spots. The electron beam incidence is parallel to the [001] direction, and the reflection spots are indexed using a simple hexagonal notation. The dark field image taken using the (1/2 $\bar{1}$ 0) superlattice spot (indicated by a red circle in Fig. S2-1a) clearly exhibits the presence of dark-line contrasts (Fig. S2-1b), while there exists no line contrast in the corresponding dark-field image taken using the (0 $\bar{1}$ 0) fundamental reflection spot (indicated by a blue circle in Fig. S2-1a) as shown in Fig. S2-1c. Similar features can be observed in Fe$_{0.43}$: the superlattice dark field image (Fig. S2-2b) taken using the ($\bar{1}$/3 $\bar{1}$/3 0) spot (indicated by a red circle in Fig. S2-2a) clearly shows dark-line contrasts, while no contrast exists in the corresponding fundamental dark-field image (Fig. S2-2c) taken using the ($\bar{1}$ 0 0) spot (indicated by a blue circle in Fig. S2-2a). It has been well established that dark-field images taken using superlattice reflection spots can exhibit diffraction contrasts for antiphase boundaries, while fundamental dark-field images do not[3]. These results indicate that the dark-line contrasts in the superlattice dark-field images in both Fe$_{0.25}$ and Fe$_{0.43}$ are due to the presence of antiphase domain boundaries.

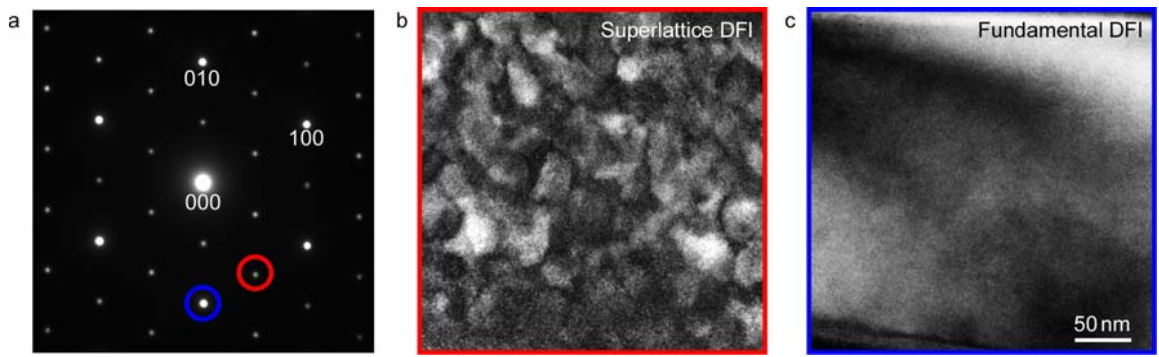

Fig. S2-1. **a,** The electron diffraction pattern of Fe$_{0.25}$. The electron beam incidence is parallel to the [001] direction, and the reflection spots are indexed using a simple hexagonal notation. **b,** The dark field images

S3

taken using the (1/2 $\bar{1}$ 0) superlattice spot. **c,** The dark-field image taken using the (0 $\bar{1}$ 0) fundamental spot. The (1/2 $\bar{1}$ 0) superlattice and (0 $\bar{1}$ 0) fundamental spots are indicated by red and blue circles in **a**, respectively, and **b** and **c** are obtained from the identical area.

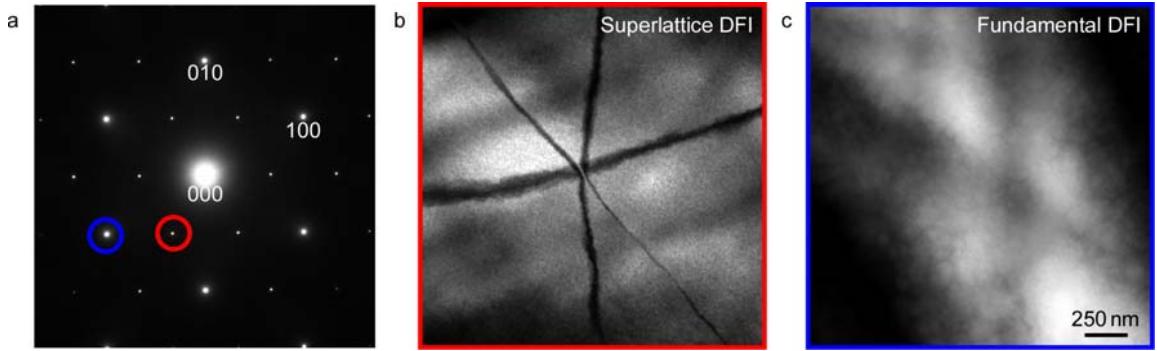

Fig. S2-2. **a,** The electron diffraction pattern of Fe$_{0.43}$. The electron beam incidence is parallel to the [001] direction, and the reflection spots are indexed using a simple hexagonal notation. **b,** The dark field images taken using the ($\bar{1}$/3 $\bar{1}$/3 0) superlattice spot. **c,** The dark-field image taken using the ($\bar{1}$ 0 0) fundamental reflection spot. The ($\bar{1}$/3 $\bar{1}$/3 0) superlattice and ($\bar{1}$ 0 0) fundamental spots are indicated by red and blue circles in **a**, respectively, and **b** and **c** are obtained from the identical area.

**Section S3. Extinction rule at antiphase domain boundaries and identification of antiphase domains in Fe$_{1/4}$TaS$_2$**

Figures S3a-S3c show a set of dark-field images taken using the superlattice reflection spots indicated by S1, S2, and S3 in the electron diffraction pattern shown in Fig. 1a, respectively. The antiphase boundaries can be observed as dark-line contrasts in the large area of the images. There exists an extinction rule for antiphase boundaries. For example, the antiphase boundary indicated by white arrows appears in the dark-field images taken using the S1= ($\bar{1}$/2 0 0) (Fig S3a) and S2=(0 1/2 0) (Fig. S3b) superlattice spots, but it disappears in the S3=(1/2 $\bar{1}$/2 0) dark-field image (Fig. S3c). This extinction rule allows us to identify the four different antiphase domains in a complicated domain configuration. See Section S5 for the relationship between the extinction rule and local structure at antiphase boundaries in Fe$_{1/4}$TaS$_2$.

S4

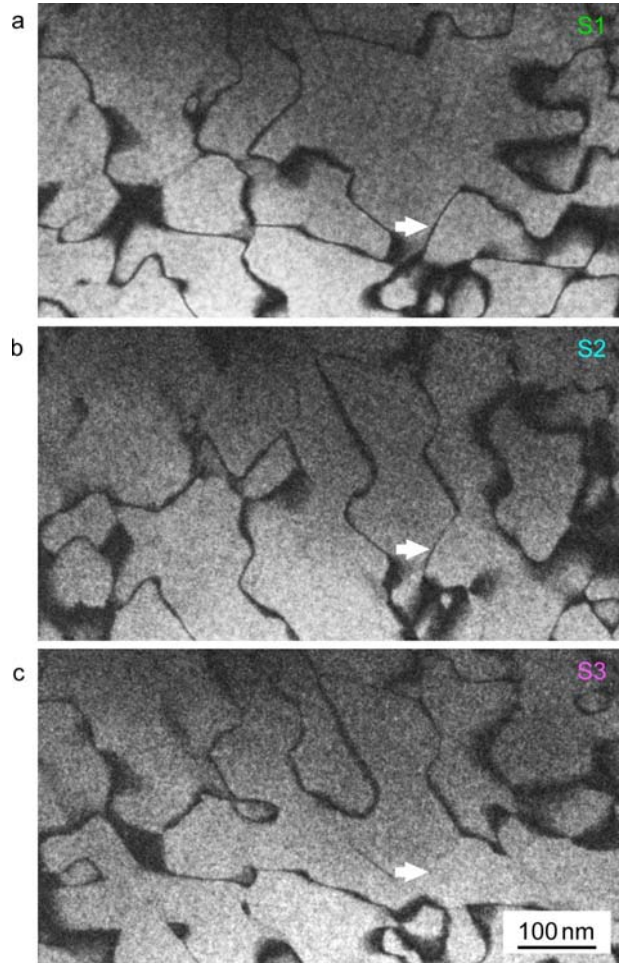

Fig. S3. The superlattice dark-field images of Fe$_{0.25}$. **a**, **b**, and **c** are taken using the superlattice reflection spots indicated by S1, S2, and S3 in the electron diffraction pattern shown in Fig. 1**a**, respectively.

**Section S4. Local structures at antiphase domain boundaries in Fe$_{1/4}$TaS$_2$ and Fe$_{1/3}$TaS$_2$**

A model of the local structure near an antiphase domain boundary in Fe$_{1/4}$TaS$_2$ is shown in Fig. S4a. Both red and blue spheres depict the Fe ions in two neighboring antiphase domains, and the antiphase boundary is highlighted with a yellow band. Antiphase shifts occur parallel to the S1 and S2 directions (perpendicular to the superlattice fringes in Fig. 2d and 2e), while no antiphase shift can be seen along the S3 direction (perpendicular to the superlattice fringes in Fig. 2f). Figure S4b exhibits a local structural model near the simultaneous antiphase and chiral domain boundary between AC- and BC-type domains in Fe$_{1/3}$TaS$_2$. Both red and green spheres depict the Fe ions in the upper layer, while small blue spheres indicate the Fe ions in the lower layer. The simultaneous antiphase and chiral domain boundary is highlighted with a yellow band. The Fe ions in the upper layer (red and green spheres) exhibit the antiphase relationship between 2D A- and B-type supercells, while those in the lower layer (small blue spheres) maintain a C-type supercell across the boundary, resulting in the boundary between AC- and BC-type domains. In contrast to $Fe_{1/4}TaS_2$, the antiphase shifts exist along all three equivalent $\vec{q}$ directions (i.e., ($\bar{1}/3\ \bar{1}/3\ 0$), ($2/3\ \bar{1}/3\ 0$), and ($\bar{2}/3\ 1/3\ 0$) directions) in $Fe_{1/3}TaS_2$. This indicates that the domain boundaries of the $\sqrt{3}a \times \sqrt{3}a$ superstructure are always visible in the superlattice dark-field images taken using different superlattice modulation wave vectors, which is consistent with our experimental results.

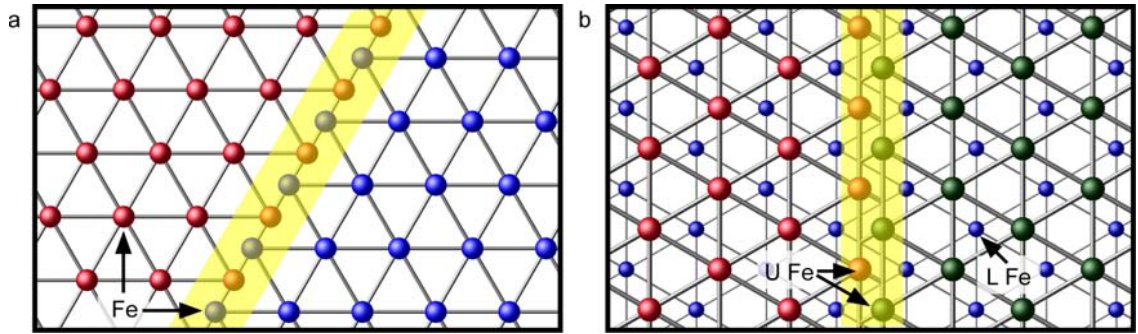

Figs. S4. Local structures at domain boundaries in **a** $Fe_{1/4}TaS_2$ and **b** $Fe_{1/3}TaS_2$. All red, blue, dark-green spheres represent Fe ions. The large and small spheres in **b** correspond to the Fe ions locating in the upper and lower Fe layers in $Fe_{1/3}TaS_2$, respectively. The domain boundaries are highlighted with yellow bands. Note that **a** denotes an antiphase boundary in $Fe_{1/4}TaS_2$, while **b** exhibits a simultaneous antiphase and chiral boundary in $Fe_{1/3}TaS_2$.

**Section S5. $Z_2 \times Z_3$ coloring of a large-range antiphase domain image of $Fe_{1/3}TaS_2$**

Figure S5 shows a large-range superlattice dark-filed image obtained on $Fe_{0.43}$. The image is taken using the ($\bar{1}/3\ \bar{1}/3\ 0$) superlattice spot denoted by S4 in Fig. 1e. The image in the entire area is a $Z_2 \times Z_3$ domain pattern. Every face (domain) in the image is surrounded by an "even" number of vertices connected by edges (domain walls), and six edges always merge at each vertex, which indicates that the antiphase domain pattern in Fig. S5 is a "6-valent graph with even-gons". In Fig. S5b, dark domains are not touching bright domains through edges, and vice versa. In addition, for example, each dark red domain is not surrounded by any light red domains. These two step proper coloring is the essence of $Z_2 \times Z_3$ coloring.

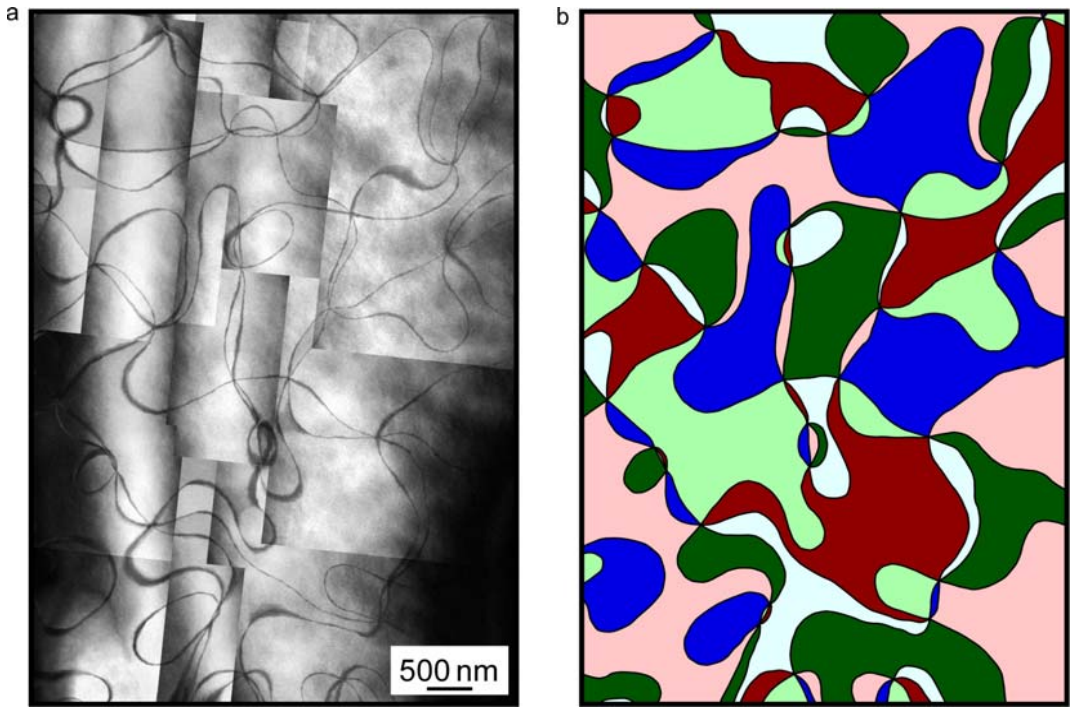

Figs. S5 **a,** A large-range superlattice dark-field image of $Fe_{0.43}$. This image is a mosaic of many dark-field images taken using the ($\bar{1}/3$ $\bar{1}/3$ 0) superlattice spot denoted by S4 in Fig. 1**e**. **b,** Cartoon of **a** with $Z_2 \times Z_3$ coloring. This cartoon corresponds to a 6-valent graph with even-gons.

**Section S6. Chirality of $Fe_{1/3}TaS_2$**

The mirror image of the crystal structure of $Fe_{1/3}TaS_2$ cannot be overlapped with the original image after any crystallographic symmetry operations, indicating the presence of structural chirality. Figures S6 show the images of a pair of $Ta-S_6$ prisms around an intercalated Fe ion in $Fe_{1/3}TaS_2$; the original image, the one after two consecutive 180º rotations around a- and c- axes, and the one after a mirror operation. The image after two consecutive 180º rotations is identical with the original one (Fig. S6a). However, the mirror image is different from either of them, and cannot be, in fact, overlapped with an image with any rotations and/or translations (Fig. S6b).

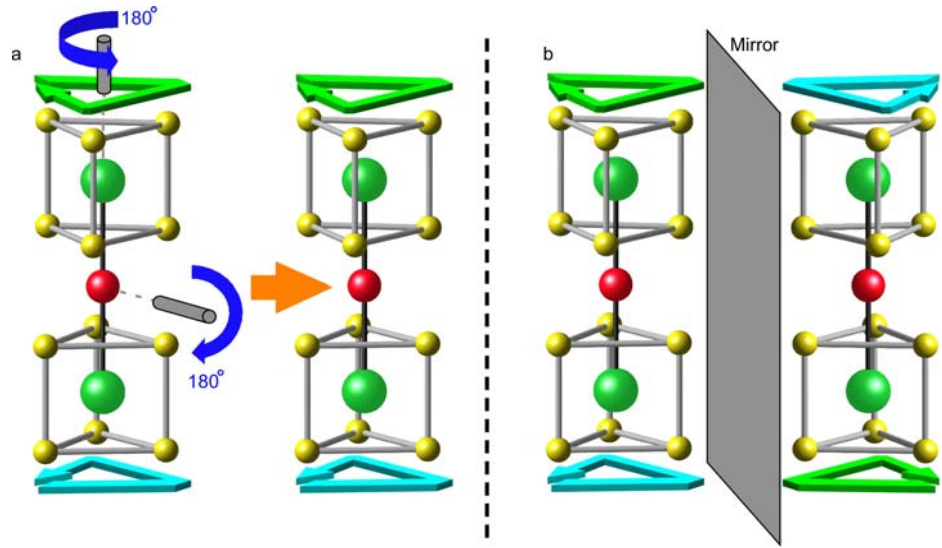

Fig. S6. Transformations of a pair of Ta-S$_6$ prisms above and below an intercalated Fe ion in Fe$_{1/3}$TaS$_2$ under various crystallographic operations. **a,** A pair of Ta-S$_6$ prisms before and after two successive 180º rotations around a- and c- axes: they are identical. **b,** A pair of Ta-S$_6$ prisms before and after mirror reflection. The mirror image cannot be overlapped with the Ta-S$_6$ pair image with any rotations and/or translations.

**Section S7. Local distortions near a vortex core with the AB/CB/CA/BA/BC/AC domain configuration**

Figure S7 displays the local lattice distortions near a vortex core with the AB/CB/CA/BA/BC/AC=C+/A-/B+/C-/A+/B- domain configuration. The A+(=BC), B+(=CA), and C+(=AB) domains have a chirality with the counterclockwise rotation of Ta-S$_6$ prisms, while those with A-(=CB), B-(=AC), and C-(=BA) domains exhibit the opposite chirality.

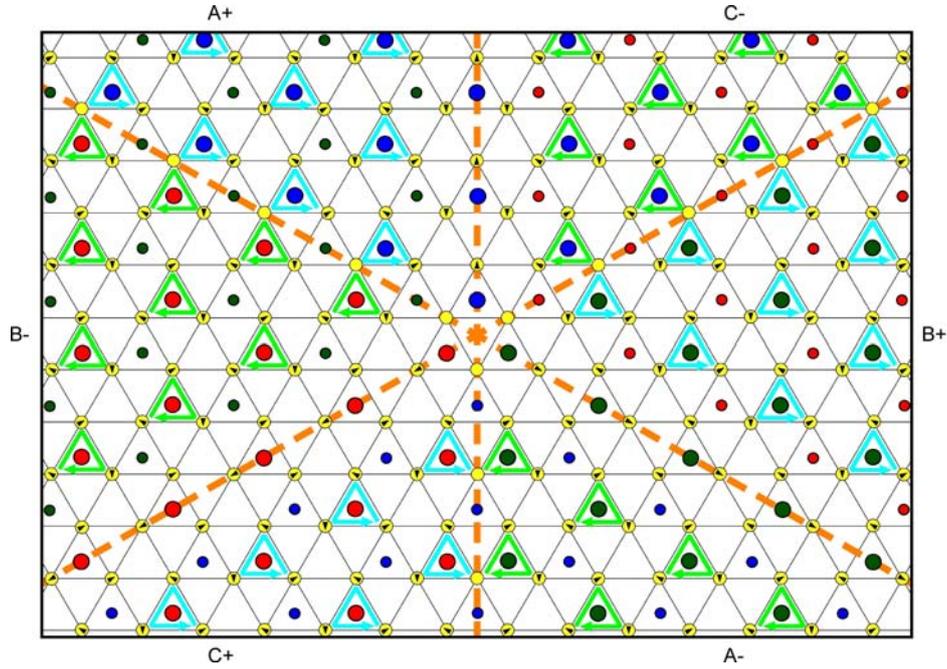

Fig. S7. Local lattice distortions near a vortex core with the A+/B-/C+/A-/B+/C- domain configuration. The large (small) red, blue, and green circles represent the Fe ions in A, B, and C type antiphase domains in the upper (lower) layer, respectively. The yellow circles show S ions, where black arrows depict the ionic displacement directions. The light blue (light green) allows correspond the counterclockwise (clockwise) rotation of Ta-$S_6$ prisms locating below the intercalated Fe ions in the upper Fe layer.

**Section S8. Difference of $Z_4$ and $Z_2 \times Z_3$ antiphase domains in size**

The $Z_2 \times Z_3$ antiphase domains tend to be significantly larger in size, compared with the $Z_4$ domains. Figures S8a and S8b show the superlattice dark field image with the electron incidence close to the [001] zone axis and the Friedel's-Pair-Breaking dark-field image obtained in Fe$_{0.34}$, respectively, taken using the ($\bar{1}/3$ $\bar{1}/3$ 0) superlattice reflection spot (indicated by arrow in the inset). In these images, the $Z_2 \times Z_3$ domains can be observed with the typical size of ~500 nm. Note that the size of the $Z_2 \times Z_3$ antiphase domains in Fe$_{0.43}$ (Fig. 1g) is ~3 μm. Remarkably, the size of the $Z_4$ antiphase domains (about 150 nm in Fe$_{0.25}$) is still smaller than that of the small $Z_2 \times Z_3$ domains in Fe$_{0.34}$, implying that the domain evolution can be related to the domain topologies. Since a $Z_4$ antiphase domain neighbors three other types of antiphase domains, it may cost more energy when antiphase boundaries move during domain growth. On the other hand, a $Z_2 \times Z_3$ domain is surrounded by only two types of antiphase-chiral domains (i.e., AB- domain can neighbor only AC- and CB-domains), probably leading to facile motion of antiphase-chiral domain boundaries that result in large domain sizes. Note that the annihilation of vortices and antivortices observed only in the $Z_2 \times Z_3$ domains may also play a crucial role in the fast growth of $Z_2 \times Z_3$ domains. It is also worthy to mention that $Fe_{1/3+\delta}TaS_2$ with more excess Fe ions tends to have larger domains. These excess Fe ions are supposed to locate unoccupied sites between ordered Fe ions consisting of the √3a×√3a superstructure. The $Fe_{1/3}TaS_2$ with more excess Fe ions can, therefore, reconstruct the √3a×√3a superstructure readily during the growth of antiphase domains. This easy reconstruction of the √3a×√3a superstructure combined with the facile boundary motion probably leads to larger antiphase domains in $Fe_{1/3+\delta}TaS_2$ with large δ. Another important fact is that no individual antiphase or chiral boundaries are observed in $Fe_{1/3}TaS_2$, indicating that the antiphase boundaries and chiral domain boundaries are mutually locked in a coherent manner. Note that domain topology does not depend on Fe composition for a given superstructure either 2a×2a or √3a×√3a.

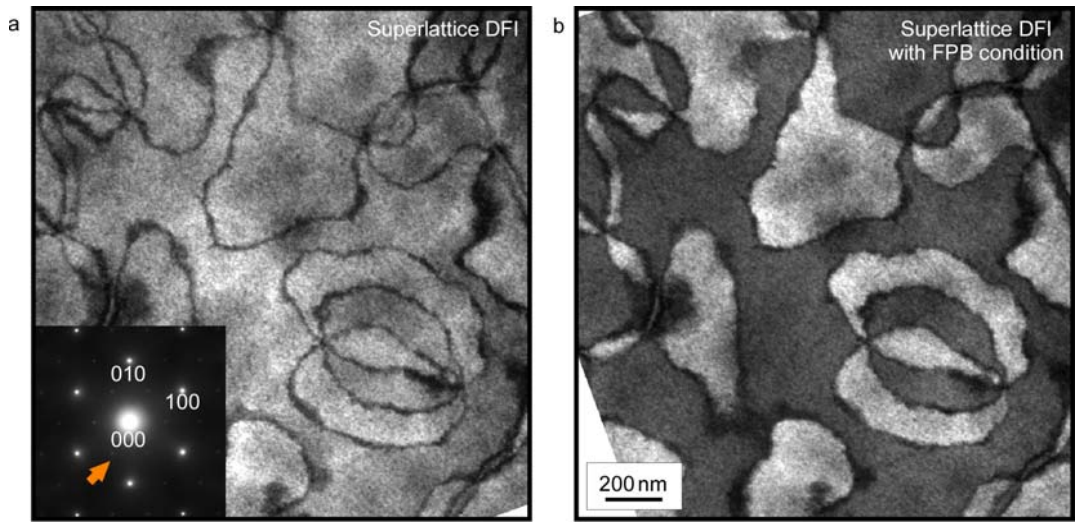

Figs. S8. The dark-field TEM images taken using the ($\bar{1}/3$ $\bar{1}/3$ 0) superlattice spot indicated by arrow in the inset of the [001] electron diffraction pattern of $Fe_{0.34}$. **a,b,** TEM images are taken with the electron incidence almost parallel to the [001] zone axis and under the Friedel's-Pair-Breaking condition, respectively. The antiphase and chiral (noncentrosymmetric) domains with $Z_2 \times Z_3$ coloring can be observed clearly with the typical domain size of ~500 nm.